\documentclass[aps,prl,superscriptaddress,twocolumn]{revtex4-1}
\usepackage{amssymb}
\usepackage{graphicx}
\usepackage{amsmath}
\usepackage{natbib}
\usepackage{siunitx}
\usepackage{textcomp}
\usepackage{bm}

\usepackage{color}
\usepackage[colorlinks]{hyperref}
\hypersetup{
    colorlinks=true,
    linkcolor=red,
    filecolor=magenta,      
    urlcolor=blue,
}

\def\bk{{\bf k}}

\def\br{{\bf r}}
\def\bj{{\bf j}}
\def\bv{{\bm v}}

\def\calD{\mathcal{D}}

\def\calP{\mathcal{P}}

\def\e{\epsilon}

\def\pa{\partial}
\def\nn{\nonumber}

\begin{document}

\title{Dynamical Kosterlitz-Thouless Theory for Two-Dimensional Ultracold Atomic Gases}
\author{Zhigang Wu}
\affiliation{Guangdong Provincial Key Laboratory of Quantum Science and Engineering, Shenzhen Institute for Quantum Science and Engineering, Southern University of Science and Technology, Shenzhen 518055, Guangdong, China}
\author{Shizhong Zhang}
\affiliation{Department of Physics and HKU-UCAS Joint Institute for Theoretical and Computational Physics at Hong Kong, The University of Hong Kong, Hong Kong, China}
\author{Hui Zhai}
\affiliation{Institute for Advanced Study, Tsinghua University, Beijing,
100084, China}
\affiliation{Center for Quantum Computing, Peng Cheng
Laboratory, Shenzhen 518055, China}

\date{\today }

\begin{abstract} 
In this letter we develop a theory for the first and second sound in a two-dimensional atomic superfluid across the superfluid transition based on the dynamical Koterlitz-Thouless theory. We employ a set of modified two-fluid hydrodynamic equations which incorporate the dynamics of the quantised vortices, rather than the conventional ones for a three-dimensional superfluid. As far as the sound dispersion equation is concerned, the modification is essentially equivalent to replacing the static superfluid density with a frequency dependent one, renormalised by the frequency dependent ``dielectric constant" of the vortices. This theory has two direct consequences. First, because the renormalised superfluid density at finite frequencies does not display discontinuity across the superfluid transition, in contrast to the static superfluid density, the sound velocities vary smoothly across the transition. Second, the theory includes dissipation due to free vortices, and thus naturally describes the sound-to-diffusion crossover for the second sound in the normal phase. With only one fitting parameter, our theory gives a perfect agreement with the experimental measurements of sound velocities across the transition, as well as the quality factor in the vicinity of the transition. The predictions from this theory can be further verified by future experiments.  
\end{abstract}
\maketitle

Topological defects, such as quantised vortices in superfluids, are of fundamental importance in physics of many two-dimensional (2D) systems~\cite{Mermin}. Phase transitions driven by these defects, known as the Berezinskii-Kosterlitz-Thouless (BKT) transitions~\cite{Berezinsky,Kosterlitz_1972,Kosterlitz_1973,Jose,Kosterlitz_review}, are found to exist in a wide range of 2D systems~\cite{Kosterlitz_review}, including Helium films~\cite{Bishop_I,Bishop_II}, superconducting films~\cite{Hebard}  and 2D ultracold atomic gases~\cite{Hadzibabic,Murthy}. Although conceptually similar to the other two kinds of systems, the study of ultracold atomic gases can in fact significantly enrich the BKT physics~\cite{Hadzibabic2011}. From the perspective of experimental technique, for example, matter wave interferometry allows visualisation of quantised vortices and thus potentially direct observation of the proliferation of free vortices across the BKT transition~\cite{Hadzibabic2011}. 

Nevertheless, a more important conceptual aspect is the study of sound propagation, which behaves very differently in these systems. In the superconducting films, the sound wave corresponds to a plasma mode with the square-root dispersion due to the Coulomb interaction between charged electrons~\cite{Mishonov,Buisson}. As for the Helium films, because they form on a substrate which clamps the motion of the normal component, and because they are almost incompressible, the only type of sound allowed is a surface wave of the superfluid component known as the third sound~\cite{Atkins1,Bergman_I,Kagiwada}. The 2D atomic gas, on the other hand, is charge neutral compared to superconducting films; it is also isolated from any other environment and is much more compressible compared to Helium films,  which permits motions of both the normal and superfluid components. Thus, the cold atomic gas provides a platform for the experimental exploration of both the first and the second sound in a 2D system across the BKT transition. 

Indeed, such an experiment has been carried out recently in a ultracold 2D Bose gas~\cite{Ville_2018}. However, a discrepancy is found between the experimental observation and the theories based on the Landau two-fluid theory~\cite{Ozawa_2014,Liu2014,Ota_2018,Ville_2018}. Since the static superfluid density has a discontinuous jump across the BKT transition, the standard Landau two-fluid theory naturally predicts a discontinuity of the second sound velocity across the BKT transition. However, such a discontinuity is not found in the experiment. Instead, the experiment finds a second sound mode with a smoothly varying velocity and a rapidly increasing damping rate across the BKT transition. Although several theoretical works offer possible explanations with various numerical approaches~\cite{Ota_PRL_2018,Cappellaro,Singh2020}, most of them assume that the experimental system is in the collisionless regime rather than the hydrodynamic regime. Thus it remains an open question as to whether the sound propagation in the actual physical system, whose collision rate is several times larger than the probed sound frequencies~\cite{Ville_2018}, can still be understood in term of  hydrodynamics.  

This discrepancy thus calls for a serious revisit of the two-fluid hydrodynamics for 2D BKT superfluids. In fact, earlier pioneering works have developed a so-called dynamical Kosterlitz-Thouless (KT) theory~\cite{Ambegaokar1,Ambegaokar2,Ambegaokar3,Minnhagen}  for understanding the third sound propagation in Helium films, which shows that a proper treatment of the vortex dynamics is crucial, especially in the vicinity of the superfluid transition. However, such a contribution from vortex dynamics is not contained in the standard Landau two-fluid hydrodynamic theory~\cite{Pitaevskii}. Because the contribution of vortex dynamics is much more significant in 2D than in 3D, this explains why Landau two-fluid theory works well in 3D but fails in 2D. So far, a dynamical Kosterlitz-Thouless theory for the first and second sound propagation in  2D atomic gases is still lacking. 

In this letter we present such a theory, and we show that the experimental findings can be well explained within the framework of hydrodynamic theory when the vortex dynamics is properly taken into account. The key result of our theory is a new dispersion equation that determines the velocity and damping of both branches of the sound.  Based on this equation, we show that the both sound velocities vary smoothly across the BKT transition. For the second sound in particular, we  also show that the damping increases rapidly above the BKT transition due to the proliferation of free vortices, eventually turning the wave propagation into a diffusive mode. 
With only one fitting parameter, our theory  agrees quantitatively with experimental measurements of a set of second sound velocities across the BKT transition, as well as the quality factor in the vicinity of the transition.

\textit{Dynamical Kosterlitz-Thouless Theory.} We begin by deriving a sound dispersion equation for 2D atomic superfluids from a set of five basic hydrodynamic equations. The first three of them are essentially the conservation laws of mass, momentum and entropy, i.e., 
\begin{align}
\frac{\partial \rho}{\partial t} &= - {\bm \nabla}\cdot {\bm j} \label{mc} \\
\frac{\partial {\bm j}}{\partial t}& =- {\bm \nabla} P \label{Euler} \\
\frac{\partial (\rho s)}{\partial t} & =  -{\bm \nabla}\cdot(\rho s {\bm v}_n) ,\label{ec}
\end{align}
where $\rho$ is the mass density, ${\bm j}$ is the mass current, $P$ is the pressure and $s$ is the entropy per unit mass. In terms of the  ``bare" superfluid density $\rho_s^{0}$ and the ``bare" normal fluid density $\rho_n^0$, we can write 
\begin{equation*}
\rho = \rho_s^{0}+ \rho_n^{0} \qquad {\bm j} = \rho_s^0 {\bm v}_s + \rho_n^0 {\bm v}_n,
\end{equation*}
where ${\bm v}_s$ and ${\bm v}_n$ are the superfluid and normal fluid velocity fields, respectively. These ``bare" densities contain only effects of short wavelength fluctuations, in contrast to the renormalised ones introduced later in Eq. (\ref{rhosomega})-(\ref{rhosomega2}). The three equations (\ref{mc}-\ref{ec}) are the same as those in the Landau two-fluid hydrodynamics. 

The fourth equation is the equation of motion for the superfluid component~\cite{Ambegaokar1}
\begin{align}
\frac{\partial {\bm v}_s}{\partial t} + \hat z \times {\bm J}_{v} = - {\bm \nabla} \mu,
\label{sequ}
\end{align}
where $\mu$ is the local chemical potential and $ {\bm J}_{v}$ is the current of the quantised vortices to be defined shortly. The second term in Eq.~(\ref{sequ}), absent in the corresponding equation for the 3D superfluids, accounts for the contribution of the quantised vortices in 2D superfluids. The density of quantised vortices can be written as
$
N({\bm r},t) = ({2\pi \hbar}/{m})\sum_i n_i \delta[{\bf r}- {\bf r}_i(t)]
$,
where ${\bf r}_i$ is the position of the $i$-th vortex and $n_i = \pm 1$ describes the direction of circulation for this vortex.   
Then ${\bm J}_{v} ({\bf r},t)$ can be defined as 
$
{\bm J}_{v} ({\bf r},t) = ({2\pi \hbar}/{m})\sum_i n_i ({d{\bf r}_i}/{dt})\delta[{\bf r}- {\bf r}_i(t)]
$,
such that the vortices obey the equation of continuity 
$
{\partial N({\bf r},t)}/{\partial t} =  -{\bm \nabla}\cdot {\bm J}_{v}
$. By viewing vortices as charged particles and drawing on an analogy to a 2D plasma,  it can be shown that ${\bf J}_{v} ({\bf r},t)$ is related to the relative velocity fields of the superfluid and normal fluid component via an ``Ohm's law"~\cite{Ambegaokar1,Ambegaokar3}
\begin{equation}
{\bf J}_{v} (\br,t) = \int dt'  \sigma(t-t') \hat z \times \left [{\bm v}_n(\br,t') - {\bm v}_s(\br,t') \right ], 
\label{Ohmslaw}
\end{equation}
where $ \sigma$ is a complex ``conductivity" for the vortices under the ``electric field" $\hat z \times \left ( {\bm v}_n - {\bm v}_s \right )$. In the frequency space, $\sigma$ can be written in terms of the dynamical ``dielectric constant" $\epsilon(\omega)$ as
\begin{align}
 \sigma(\omega) = -i\omega [\epsilon(\omega) - 1].
\end{align} 

Equations (\ref{mc}-\ref{Ohmslaw}) form a complete set of basic hydrodynamic equations for the 2D superfluid. Considering small deviations of relevant physical quantities from their equilibrium values and following standard derivations, we  arrive at the following sound dispersion equation~\cite{SM}
\begin{equation}
\left [ \frac{\omega^2}{k^2} - \frac{1}{\rho \kappa_T}\right ]\left [\frac{\omega^2}{k^2} - \frac{Ts^2\rho_s(\omega)}{c_v \rho_n(\omega)} \right ]- \frac{1}{\rho}\left(\frac{1}{\kappa_s} - \frac{1}{\kappa_T} \right )\frac{\omega^2} {k^2} = 0,
\label{sound}
\end{equation}
where  $\kappa_T = \rho^{-1}\left( {\partial \rho}/{\partial P}\right )_T$ and $\kappa_s = \rho^{-1}\left( {\partial \rho}/{\partial P}\right )_s$  are the isothermal and isoentropic compressibility respectively, and $c_v = T (\partial s/ \partial T)_\rho $ is the specific heat at constant volume. Here we introduce the frequency dependent superfluid and normal density as 
\begin{align}
\rho_s(\omega) &=  \frac{\rho_s^0}{\epsilon(\omega)},\label{rhosomega}\\
\rho_n(\omega) &= \rho - \rho_s(\omega).\label{rhosomega2}
\end{align}
Note that the physical densities $\rho$, $\rho_s^0$ and the entropy $s$ entering Eqs.~(\ref{sound})-(\ref{rhosomega2}) all take their equilibrium values. 

Equation (\ref{sound}), which determines the dispersions of the sound propagation in the 2D superfluid, represents the central result of this paper. Remarkably, despite the modification of one hydrodynamic equation and the addition of another, the resulting sound dispersion equation  (\ref{sound}) has nearly an identical form as its 3D superfluid counterpart~\cite{Pitaevskii}, the only difference being that the static superfluid and normal density are now replaced by the frequency-dependent densities defined in Eqs.~(\ref{rhosomega})-(\ref{rhosomega2}). In other words, by explicitly including the vortex dynamics in the hydrodynamic equations, the effect on sound dispersion is equivalent to renormalizing the bare superfluid density by the frequency-dependent dielectric constant of quantised vortices. In fact, by setting the dynamical dielectric constant $\epsilon(\omega) = 1$ in the absence of vortices, Eq.~(\ref{sound}) immediately reduces to the familiar sound dispersion equation in the 3D Landau two-fluid theory. For 2D superfluid, $\epsilon(\omega)$ is generally complex, meaning that the presence of the quantised vortices in such systems not only modifies the sound velocity but also induces the damping.

\begin{figure}[t]
\centering
\includegraphics[width=1\linewidth]{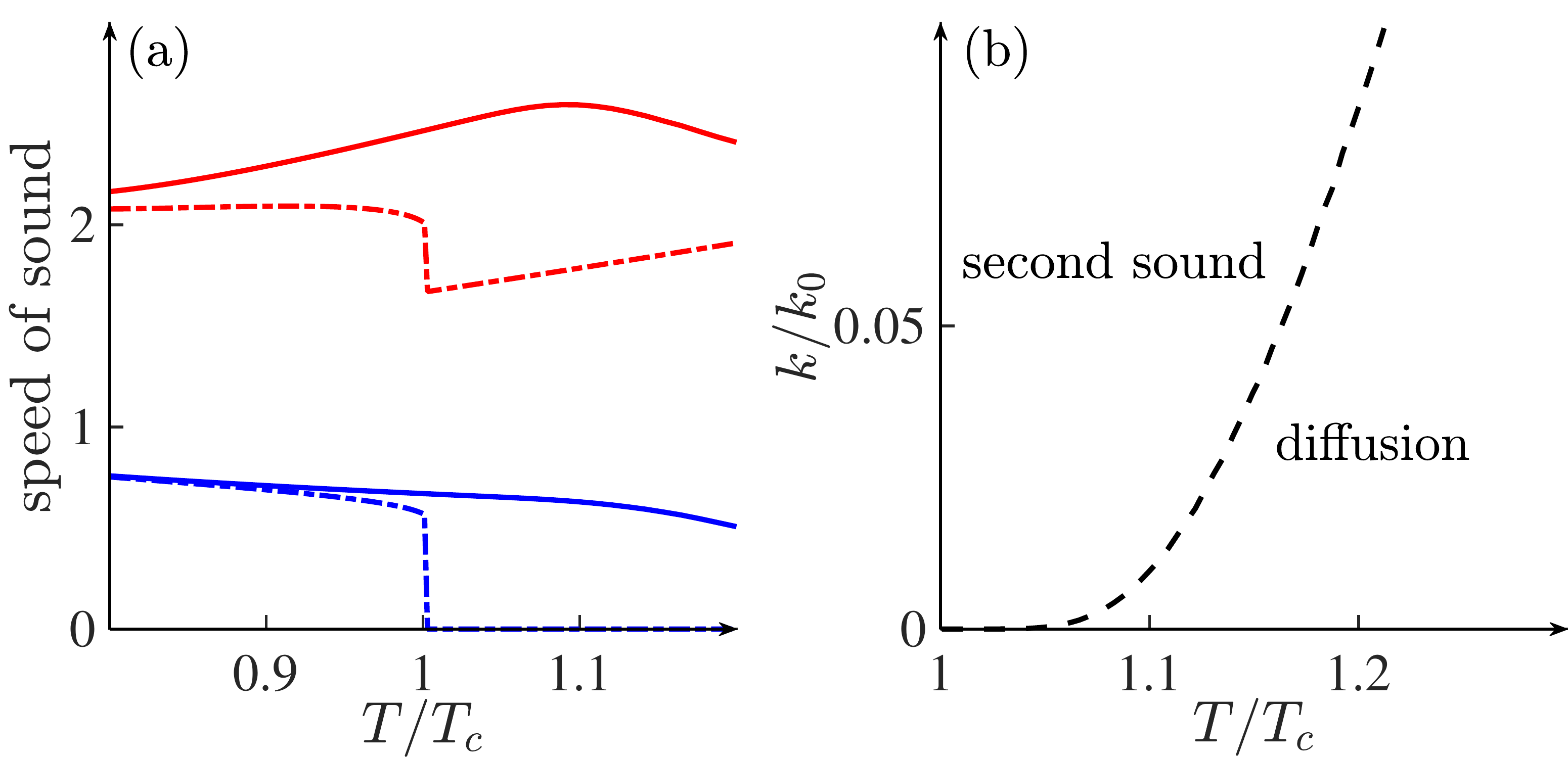}
\caption{(a) Speed of the first (upper curves) and second sound (lower curves) as a function of temperature for a 2D superfluid of weakly interacting bosons across the BKT transition: our dynamical KT theory (solid lines) vs. the Landau two-fluid theory (dashed lines).  (b) A sound-to-diffusion crossover in the $k-T$ plane for the second sound, predicted by our theory for a 2D superfluid of weakly interacting bosons above the BKT transition. Here $k_0$ is a basic unit of the wavevector. }
\label{prediction}
\end{figure}

\textit{General Features of the Sound Modes.}
Before proceeding to more detailed calculations, we first discuss two general features of the sound modes predicted by the dynamical Kosterlitz-Thouless theory, which can be inferred from the general properties of $\epsilon(\omega)$ and Eq.~(\ref{sound}). For this purpose, we  express the solutions to Eq.~(\ref{sound}) as $\omega_\alpha(k) = \varpi_\alpha (k) + i\gamma_\alpha (k)$, where $\varpi_\alpha (k) $ and $\gamma_\alpha (k)$ are the real and imaginary parts of the dispersion respectively, and  $\alpha = 1,2$ denotes the two sound branches. As in the experiment~\cite{Ville_2018}, the velocity of sound here is defined as $c_\alpha = \varpi_\alpha (k) / k$.  

\textbf {(i) Continuity of Sound Velocities.} Here we should first emphasize a difference between $\epsilon(\omega)$ at zero frequency $\epsilon(\omega=0)$ and $\epsilon(\omega)$ at finite frequencies. At $\omega =0$, $\epsilon(\omega=0)$ is real and finite for $T<T_\text{c}$ but immediately diverges at $T=T^{+}_\text{c}$, where $T_c$ is the BKT transition temperature. Hence, the non-analyticity of $\epsilon(\omega=0)$ as a function of the temperature leads to a finite $\rho_\text{s}(\omega=0)$ at $T^{-}_\text{c}$ but a vanishing $\rho_\text{s}(\omega=0)$ at $T^+_\text{c}$. Since the conventional  Laudau two-fluid theory uses the static superfluid density $\rho_s(\omega=0)$ in the hydrodynamic equations, it predicts discontinuity in sound velocities as a result of the sudden jump of $\rho_s(\omega = 0)$ at $T_c$~\cite{Ozawa_2014,Ota_2018} . If $\kappa_\text{s}=\kappa_\text{T}$, the density and temperature fluctuations are decoupled and the discontinuity exists only in the second sound, i.e., the temperature wave. Generally, $\kappa_\text{s}\neq \kappa_\text{T}$, so the aforementioned two types of fluctuations are coupled and the discontinuity exists in both two sound modes, as shown in Fig. \ref{prediction}(a) by the dashed lines.  For any finite $\omega$, however, $\epsilon(\omega)$ is always finite and is a smooth function of $T$ across BKT transition. This leads to smooth sound velocities in the first and the second sound. We calculate the sound velocities from  Eq.~(\ref{sound}) using actual parameters of a weakly interacting 2D Bose gas, and the results are shown by solid lines in Fig. \ref{prediction}(a).   

\textbf{(ii) Sound-to-Diffusion Crossover.} The second general feature our theory predicts is the second sound to diffusion crossover as a result of the proliferation of free vortices above $T_c$. In the temperature regime slightly above $T_c$, we can write $\epsilon(\omega)\approx \epsilon_b+i\sigma_f/\omega$~\cite{Ambegaokar1}, where $\e_b$ is the bound vortex pair contribution and $\sigma_f$ accounts for the ``conductivity" of free vortices. In the vicinity of $T_c$, $\epsilon_b$ is a slowly varying function  and can be approximated as a constant, while $\sigma_f$ is  proportional to the the density of free vortices and increases rapidly as the temperature increases above $T_c$. For clarity, let us first demonstrate the crossover for a simple situation with $\kappa_s = \kappa_T$ in Eq.~(\ref{sound}), where the second sound corresponds to a pure temperature wave governed by the following equation
\begin{align}
\omega^2 - 2 i \gamma_{2} \omega - u_{2}^2 k^2 = 0,
\label{sound_f}
\end{align}
where $\gamma_{2} \equiv -\frac{\rho \sigma_f }{2(\rho\epsilon_{b}-\rho^0_{s})}$ and $u_{2}  \equiv \sqrt{ \frac{Ts^2\rho^0_{s}}{c_v (\rho\epsilon_{b}-\rho^0_{s})}}$. The solution of this equation can be easily obtained as
\begin{align}
\omega_2(k) = \sqrt{u_2^2k^2 - \gamma_2^2}+ i\gamma_2.
\label{sound_fs}
\end{align}
 From Eq.~(\ref{sound_fs}) we can immediately define a threshold wavevector 
$k^*(T) \equiv |\gamma_2/u_2|
$ such that the second sound is a damped sound for $k > k^*(T)$ and becomes purely diffusive for $k< k^*(T)$. In other words, we can define a $k$-dependent temperature $T^*(k)$ whereby the second sound with wave vector $k$ becomes purely diffusive for $T>T^*(k)$. Since the damping parameter $\gamma_2$, proportional to $\sigma_f(T)$, vanishes below  $T_c$, $T^*(k)$ must be greater than $T_c$ for any finite $k$. That is to say, the sound-to-diffusion crossover occurs in the normal phase. A careful analysis of the more general situation with $\kappa_s\neq \kappa_T$  reveals the same behavior for the second sound~\cite{SM}. In Fig.~\ref{prediction} (b) we draw the sound-to-diffusion boundary in the $k$-$T$ plane defined by $\varpi_2(k) = 0$, again using the actual parameters for a 2D weakly interacting Bose gas.

\textit{Experimental Comparison.} Now we turn to the full solutions to Eq.~(\ref{sound}) for $T< T^*(k)$, i.e., in a temperature region where the second sound propagates well, and compare the results with the experiment. We find that, to a very good approximation, the solutions to Eq.~(\ref{sound}) can be written as 
\begin{align}
\varpi_\alpha &= k v_\alpha (\varpi_\alpha) \label{omegaalpha} \\
{\gamma_\alpha} &= -\frac{\rho}{2{\rm Re}\rho_n(\varpi_\alpha)}\frac{{\rm Im}\epsilon(\varpi_\alpha)}{{\rm Re}\epsilon(\varpi_\alpha)}\frac{u_2(\varpi_\alpha)}{u'_2(\varpi_\alpha)} v^\prime_\alpha(\varpi_\alpha), \label{gammaalpha}
\end{align}
where $'$ denotes the derivative.
Here we have defined
\begin{align}
{v_\alpha}(\omega) &= \frac{1}{\sqrt{2}} \bigg ( u_1^2+u^2_2(\omega) + \delta\kappa  \nonumber \\
 -&(-1)^\alpha \sqrt{\left [u_1^2+u^2_2(\omega) + \delta\kappa\right ]^2-4u_1^2u_2^2(\omega)}  \bigg )^{1/2},
\end{align}
where $u_1 \equiv \sqrt {\frac{1}{\rho \kappa_T}}$, $u_2(\omega) \equiv \sqrt {(Ts^2/c_v){\rm Re}[\rho_s(\omega)/ \rho_n(\omega)]}$ and $\delta\kappa \equiv (\kappa_T - \kappa_s)/\rho \kappa_s \kappa_T$.   To solve the dispersion, we need to i) determine thermodynamic quantities, i.e.,  $s$, $\kappa_s$, $\kappa_T$, $c_v$ and $\e(\omega)$, in term of microscopic parameters;  ii) solve Eq.~(\ref{omegaalpha}) self-consistently to obtain the real part of the spectrum; and iii) substitute $\varpi_\alpha(k)$ into Eq.~(\ref{gammaalpha}) to determine the imaginary part of the spectrum. 

 \begin{table}[t]
\caption{\label{parameters} Various physical quantities (the first column) needed for solving the sound dispersion equation. The second column lists the microscopic parameters that these physical quantities depend on. The third column lists the corresponding equations for calculating these quantities given in the Supplementary Material~\cite{SM}.}
\begin{ruledtabular}
\renewcommand{\arraystretch}{1.3}
\begin{tabular}{c cc}
 Quantities & Parameters & Supplementary\\ \hline
$s,\kappa_s,\kappa_T, c_v$& $ g, \mu/T$ & (S23)-(S24)\\ 
$\rho_s^0$&$g, \mu/T$ & (S33)-(S35) \\ 
$\epsilon(\omega)$ & $g,\mu/T,l_D,\omega$ & (S37)-(S41)\\ 
\end{tabular}
\end{ruledtabular}
\end{table}

For 2D ultracold Bose gas, the thermodynamic quantities can be calculated in terms of certain universal functions which depend solely on the dimensionless interaction constant  $g$ and the ratio of chemical potential to temperature $\mu/T$~\cite{Prokof'ev_I, Prokof'ev_II}. In addition, the dielectric constant $\epsilon(\omega)$ can be evaluated via the dynamical KT theory. Although the calculations of all these quantities are rather involved, they are well documented in literature~\cite{Ambegaokar1,Ambegaokar3,Yefsah,Ozawa_2014} and the details will be relegated to the supplemental material~\cite{SM}. Here, we shall only tabulate all the parameters needed for solving the sound dispersion equation in Tab.~\ref{parameters} and refer readers interested in details of the calculation to the corresponding equations in the Supplemental Material~\cite{SM}. 

\begin{figure}[t]
\centering
\includegraphics[width=0.98\linewidth]{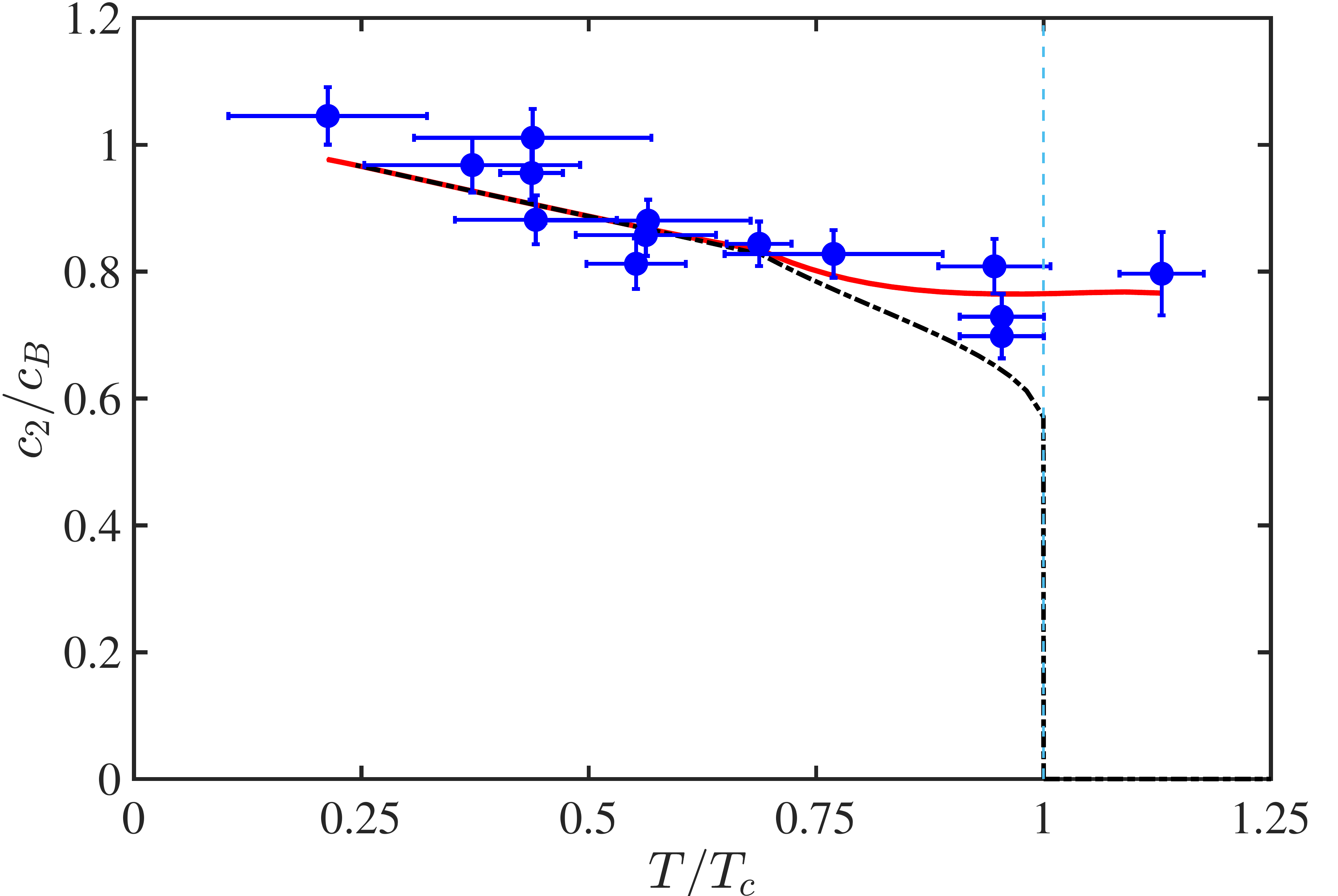}
\caption{Second sound velocity $c_2/c_\text{B}$ as a function of temperature $T/T_\text{c}$ calculated for a weakly interacting 2D $^{87}$Rb gas with $g = 0.16$. The symbols are the experimental results~\cite{Ville_2018}. The solid line is our theory with $l_D = -1.18$ and the dashed line is the Landau two-fluid theory. Here $c_\text{B} = \sqrt{g\rho}/m^{3/2}$ is the Bogoliubov sound velocity.}
\label{speed}
\end{figure}

The main parameter of the dynamical KT theory is a dimensionless quantity $l_D \equiv \ln\sqrt{2D/\omega_0 a_0^2}$, where $D$ is the vortex diffusion constant, $a_0$ is the vortex core size and $\omega_0 = c_\text{B} k_0$ is the typical frequency associated with sound propagation. Here $c_\text{B} = \sqrt{g\rho}/m^{3/2}$ is the Bogoliubov sound velocity and $k_0 = \pi/L$ is the wavevector unit for a system of size $L$ along the propagation direction. Since there is no reliable method to calculate the diffusion constant, we shall use $l_\text{D}$ as the only fitting parameter when  comparing our theoretical results to the experimental measurements. We adjust $l_\text{D}$ so as to minimize 
\begin{equation}
\delta = \sum_{i} |c_{2}(T_i) -c_{\rm exp}(T_i)|^2,
\end{equation}
where $c_{2}(T_i)$ and $c_{\rm exp}(T_i)$ are the theoretical and experimental values of the second sound velocity at temperature $T_i$, respectively. The resulting second sound velocity is shown in Fig.~\ref{speed}, where very good agreements between theory and experiment are obtained. Using $l_\text{D}$ obtained in this fitting, we find the quality factor of the second sound $Q \equiv |\varpi_2/\gamma_2|\sim 10$ at $T_c$, consistent with the experimental value of $\sim 8$. This further justifies that the value $l_\text{D}$ found from the fitting is a reasonable one. However, our theoretical values of $Q$ become significantly larger than the experimental measurements for temperatures deep in the superfluid phase. This is because the damping due to free vortices is the only damping mechanism included in this hydrodynamics theory. Deep in the superfluid phase, where the vortex contribution is insignificant, our hydrodynamics recovers the dissipationless hydrodynamics and as a result underestimates the damping.   

\textit{Concluding Remarks.} In summary, we have shown that the dynamical KT theory, which includes the dynamics of vortices, is crucial for understanding the sound propagation in 2D ultracold Bose gas. This theory essentially renormalizes the superfluid density by the complex dielectric constant, which removes the discontinuity in the superfluid density and introduces dissipation due to bound and free vortices. This leads to smooth varying sound velocities across the BTK transition and sound-to-diffusion crossover for the second sound in the normal phase. These results are consistent with the experimental findings in the weakly interacting 2D Bose gas and can be further verified by future experiments. Finally, our discussions can also be extended to two-dimensional Fermi superfluids with a BEC-BCS crossover.   

\textit{Acknowledgement}. We acknowledge helpful discussions with Zhiyuan Yao, Pengfei Zhang and Mingyuan Sun.  Z.W. is supported by NSFC (Grant No. 11974161), Key-Area Research and Development Program of Guangdong Province (Grant No. 2019B030330001) and Guangdong Provincial Key Laboratory (Grant No.2019B121203002). S.Z. is supported by HK GRF 17318316, 17305218 and CRF C6026-16W and C6005-17G, and the Croucher Foundation under the Croucher Innovation Award. H.Z.  acknowledge support by Beijing Distinguished Young Scientist Program, MOST (Grant No. 2016YFA0301600) and NSFC (Grant No. 11734010).

\bibliographystyle{apsrev4-2}

\onecolumngrid
\begin{center}
\newpage\textbf{
Supplemental Material\\[4mm]
\large Dynamical Kosterlitz-Thouless Theory for Two-Dimensional Ultracold Atomic Gases}
\\
\vspace{4mm}

{Zhigang Wu$^1$, Shizhong Zhang$^{2}$, and Hui Zhai$^{3,4}$}\\
\vspace{2mm}
{\em \small
$^1$Guangdong Provincial Key Laboratory of Quantum Science and Engineering, Shenzhen Institute for Quantum Science and Engineering, Southern University of Science and Technology, Shenzhen 518055, Guangdong, China\\
$^2$Department of Physics and HKU-UCAS Joint Institute for Theoretical and Computational Physics at Hong Kong, The University of Hong Kong, Hong Kong, China\\
$^{3}$Institute for Advanced Study, Tsinghua University, Beijing,
100084, China\\
$^4$Center for Quantum Computing, Peng Cheng
Laboratory, Shenzhen 518055, China}
\end{center}

%%%%%%%%%% Prefix a "S" to all equations, figures, tables and reset the counter %%%%%%%%%%
\setcounter{equation}{0}
\setcounter{figure}{0}
\setcounter{table}{0}
\setcounter{section}{0}
\setcounter{page}{1}
\makeatletter
\renewcommand{\theequation}{S\arabic{equation}}
\renewcommand{\thefigure}{S\arabic{figure}}
\renewcommand{\thetable}{S\arabic{table}}
\renewcommand{\thesection}{S.\arabic{section}}

\section{Derivation of the sound dispersion equation}
For 2D superfluids, the two coupled sound equations and the resulting sound dispersion equation can be derived from the following five basic equations: 
\begin{align}
\label{mcs}
\frac{\pa \rho}{\pa t} &= - \nabla\cdot \bj  \\
\label{Eulers}
\frac{\pa \bj}{\pa t} & = -  \nabla P \\
\label{ecs}
\frac{\pa (\rho s)}{\pa t}  & = - \nabla\cdot(\rho s \bv_n)  \\
\label{sequs}
\frac{\pa \bv_s}{\pa t} & = - \hat z \times {\bf J}_{vor}  - \nabla \mu \\
\label{Ohmslaws}
{\bf J}_{vor} (\br,t) &= \int  dt'  \sigma(t-t') \hat z \times \left [\bv_n(\br,t') - \bv_s(\br,t') \right ].
\end{align}
 The first hydrodynamic sound equation can be obtained from combining Eq.~(\ref{mcs}) and (\ref{Eulers}). Considering small deviations of the relevant physical quantities from their equilibrium values,  we find from Eq.~(\ref{mcs}) and (\ref{Eulers}) 
\begin{align}
\frac{\pa^2 \delta \rho}{\pa t^2} - \nabla^2 \delta P = 0,
\label{denpress}
\end{align}
where $\delta \rho$ and $\delta P$ are variations of the total density and the pressure respectively. 
The variation of the pressure can be expressed in terms of those of the density and the temperature as follows
\begin{align}
\delta P & =  \left( \frac{\pa P}{\pa \rho}\right )_T \delta \rho + \left( \frac{\pa P}{\pa T}\right )_\rho \delta T \nn \\
&= \frac{1}{\rho \kappa_T } \delta \rho + \frac{\alpha_P}{\kappa_T} \delta T ,
\end{align}
where $\kappa_T = \frac{1}{\rho}\left( \frac{\pa \rho}{\pa P}\right )_T$ is the isothermal compressibility and $\alpha_P = - \frac{1}{\rho}\left( \frac{\pa \rho}{\pa T}\right )_P$ is the isobaric thermal expansion coefficient. Substitution of the above equation into Eq.~(\ref{denpress}) yields 
\begin{align}
\frac{\pa^2 \delta \rho}{\pa t^2} - \frac{1}{\rho\kappa_T}\nabla^2 \delta \rho -\frac{\alpha_P}{\kappa_T}\nabla^2\delta T = 0.
\end{align}
In terms of the Fourier components of the density of and temperature fluctuations, we find the first hydrodynamic sound equation
\begin{align}
\left (\frac{\omega^2}{k^2}- \frac{1}{\rho\kappa_T}\right ) \delta \rho (\bk,\omega)-\frac{\alpha_P}{\kappa_T}\delta T(\bk,\omega) = 0.
\label{hydro1}
\end{align}
To obtain the second hydrodynamic sound equation, we first linearise Eq.~(\ref{ecs}) and use Eq.~(\ref{mcs}) to obtain
\begin{align}
\frac{\pa \delta s}{\pa t} = -s\frac{\rho_s^0}{\rho} \nabla\cdot(\bv_{n} - \bv_s). 
\label{ec1}
\end{align}
Using the thermodynamic relation
\begin{align}
\delta s &=\left (\frac{\pa s}{\pa T} \right )_\rho \delta T +\left ( \frac{\pa s}{\pa \rho}\right )_T \delta \rho \nn \\
& = \frac{ c_v}{T}\delta T - \frac{\alpha_P}{\rho^2 \kappa_T} \delta \rho,
\end{align}
where $c_v = (\pa s/ \pa T)_\rho/T $ is the specific heat at constant volume, Eq.~(\ref{ec1}) can be written as 
\begin{align}
 \frac{ c_v}{T}\frac{\pa \delta T}{\pa t} - \frac{\alpha_P}{\rho^2 \kappa_T} \frac{\pa \delta \rho}{\pa t} = -s\frac{\rho_s^0}{\rho} \nabla\cdot\bv_{ns},
\end{align}
where we define the relative velocity field $\bv_{ns} = \bv_n - \bv_s$. 
Fourier transforming the above equation gives 
\begin{align}
 -i\omega\left [ \frac{ c_v}{T} \delta T (\bk,\omega)- \frac{\alpha_P}{\rho^2 \kappa_T} \delta \rho(\bk,\omega)\right ] = -is\frac{\rho_s^0}{\rho} \bk\cdot\bv_{ns}(\bk,\omega).
 \label{ec2}
\end{align}
Next,  we use the Gibbs-Duhem equation
\begin{align}
\delta \mu = \frac{1}{\rho} \delta P - s \delta T,
\end{align}
and Eq.~(\ref{Eulers}) to rewrite Eq.~(\ref{sequs}) as
\begin{align}
\frac{\rho_n^0}{\rho} \frac{\pa}{\pa t} \hat z\times \bv_{ns} +  {\bf J}_{vor} = -s\hat z \times \nabla \delta T. 
\end{align}
In momentum frequency space we have
\begin{align}
\frac{\rho_n^0}{\rho} (-i\omega) \hat z\times \bv_{ns}(\bk,\omega) +  {\bf J}_{vor}(\bk,\omega) = -is\hat z \times \bk \delta T (\bk,\omega). 
\label{sequ1}
\end{align}
To close the above equation, we need Eq.~(\ref{Ohmslaws}) which, in Fourier space, is 
\begin{align}
{\bf J}_{vor}(\bk,\omega) & =  \sigma(\omega) \hat z\times \bv_{ns}(\bk,\omega) \nn \\
& =  -i\omega [\e(\omega) - 1]\hat z\times \bv_{ns}(\bk,\omega), 
\end{align}
where in the second line expressed the complex conductivity $ \sigma$ in terms of the dielectric constant.
Substituting the above expression into Eq.~(\ref{sequ1}) and multiply both sides by $\bk$ we arrive at 
\begin{align}
 -i\omega \left( \e(\omega)- \frac{\rho_s^0}{\rho} \right ) \bk\cdot \bv_{ns}(\bk,\omega) = -i s k^2 \delta T(\bk,\omega).
 \label{sequ2}
\end{align}
Now we use Eq.~(\ref{sequ2}) to eliminate the $\bk\cdot \bv_{ns}(\bk,\omega) $ term in Eq.~(\ref{ec2}) and we find the second hydrodynamic sound equation 
\begin{align}
\frac{\alpha_P}{\rho^2 \kappa_T} \frac{\omega^2}{k^2} \delta \rho(\bk,\omega) + \left [\frac {\rho_s^0}{ \rho }s^2 \frac {1}{\e(\omega) - \rho_s^0/\rho } - \frac{ c_v}{T} \frac{\omega^2}{k^2}\right ] \delta T(\bk,\omega)= 0. 
\label{hydro2}
\end{align}
Combining the two hydrodynamic sound equations (\ref{hydro1}) and (\ref{hydro2}), we finally arrive at the equation that determines the dispersion of sound propagation in 2D superfluids 
\begin{align}
\left [ \frac{\omega^2}{k^2} - \frac{1}{\rho \kappa_T}\right ]\left [\frac{\omega^2}{k^2} - \frac{Ts^2\rho_s(\omega)}{c_v \rho_n(\omega)} \right ] - \frac{1}{\rho}\left(\frac{1}{\kappa_s} - \frac{1}{\kappa_T} \right )\frac{\omega^2} {k^2} = 0,
\label{sounds}
\end{align}
where the frequency-dependent superfluid density is defined as 
\begin{align}
\rho_s(\omega) = \rho_s^0/\e(\omega)
\label{rhosomegas}
\end{align}
and the corresponding normal density is  $\rho_n(\omega) = \rho - \rho_s(\omega)$.

\section{Second sound to diffusion transition in the general case of $\kappa_s \neq \kappa_T$}
In the general case of $\kappa_s \neq \kappa_T$, we make the same approximation $\e(\omega,T) \approx \e_b(0,T_c^{-}) + i\sigma_f(T)/\omega
$ in the sound dispersion equation (\ref{sounds}) and arrive at  the following quartic equation
\begin{align}
\omega^4 -2i \gamma_2 \omega^3 -(u_1^2 +u_2^2 + \delta \kappa) k^2 \omega^2 +2i\gamma_2 (u_1^2+\delta \kappa) k^2 \omega +u_1^2 u_2^2 k^4 = 0,
\label{quartic}
\end{align}
where $u_1 \equiv  \sqrt{{1}/{\rho \kappa_T}}$, $u_2 \equiv \sqrt{ {Ts^2\rho_s(0)}/{c_v \rho_n(0)}}$, $\delta \kappa \equiv { (\kappa_T - \kappa_s)/\rho \kappa_s \kappa_T}$ and $\gamma_{2} \equiv -\frac{\rho \rho_{s}(0) }{2\rho_s^0 \rho_{n}(0)}\sigma_f$. Here we have used the fact that $\e_b(0,T_c^-) = \rho_s^0/\rho_s(0)$ (see later). All the densities and other thermodynamic quantites appearing in these velocity and damping coefficients are approximated by the values at $T_c$, since their temperature dependences are much weaker than that of $\sigma_f(T)$ in the vicinity of $T_c$. 

\begin{figure}[th]
\centering
\includegraphics[width=0.45\linewidth]{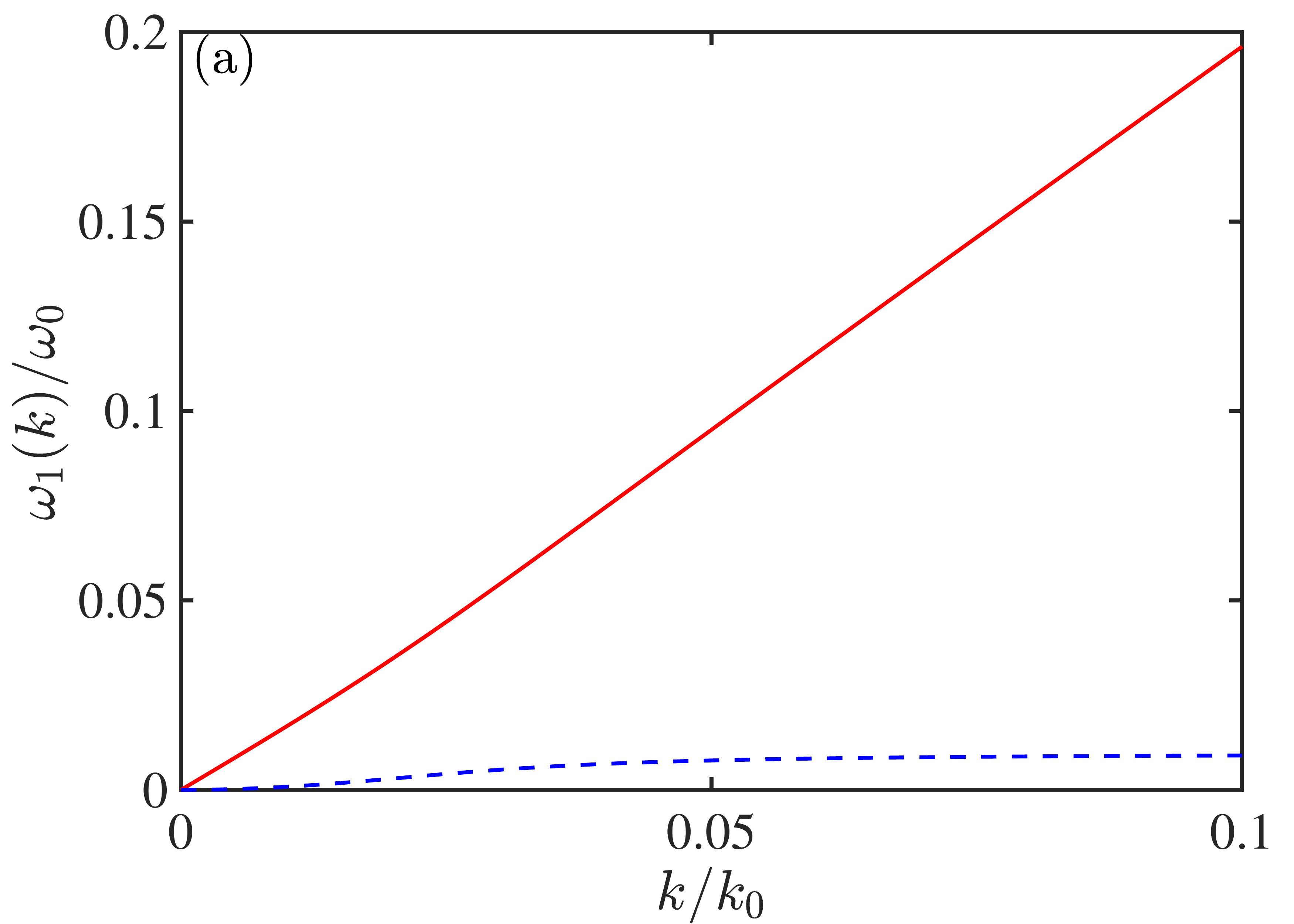}
\includegraphics[width=0.45\linewidth]{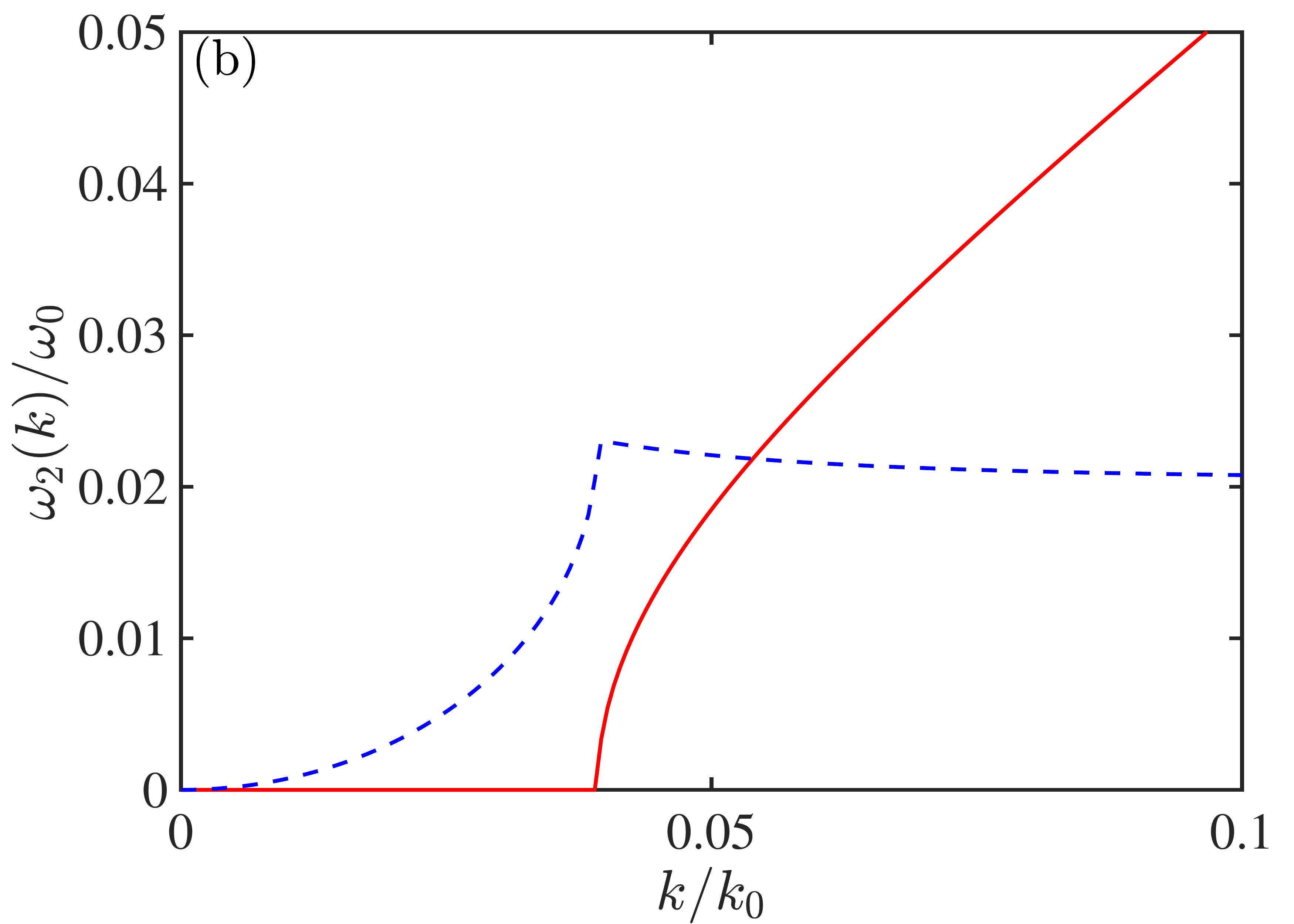}
\caption{(a) The real (solid) and imaginary (dashed) part of the first sound dispersion. (b) The real (solid) and imaginary (dashed) part of the second sound dispersion. The calculations are done for a Bose gas with $g = 0.16$, $T/T_c = 1.15$ and $l_D = -1$.}
\label{omega12}
\end{figure}

The above quartic equation can be easily solved numerically to obtain $\omega(k)$ for any specific $T$. For sufficiently large $k$, we find that the four complex solutions can be organized into two pairs, each of which has two solutions with opposite real parts. We can thus identify the two solutions with the positive real parts, one in each pair, as the dispersions for the two branches of the sound. As we decrease $k$, the real part of the solution corresponding to the second sound decreases gradually and becomes zero for $k< k^*(T)$. This is shown in Fig.~\ref{omega12}. In other words, we find purely imaginary solutions for the second sound branch for $k< k^*(T)$, signaling that the sound mode transitions into a diffusive mode. This is similar to the more special case of $\kappa_s = \kappa_T$. The $k-T$ boundary shown in the main text was obtained by solving Eq.~(\ref{quartic}) for a weakly interacting 2D Bose gas. 
\section{Calculation of various thermodynamic quantities of the 2D weakly interacting Bose gas}
For a weakly interacting 2D Bose gas, the various thermodynamic quantities, such as $s$, $\kappa_s$, $\kappa_T$ and $c_v$, can be calculated in terms of certain universal functions dependent only on the variable $x = \mu/T$ and the dimensionless coupling constant $g$. First, All these quantities can be expressed in terms of the dimensionless reduced pressure $\calP$ and the phase space density $\calD$~\cite{Ozawa_2014} 
\begin{align}
\calP(x,g) \equiv \lambda_T^2 \frac{P}{T} \qquad\qquad \calD(x,g) \equiv \lambda_T^2 \frac{\rho}{m},
\label{calPcalD}
\end{align}
where $\lambda_T = \hbar \sqrt{2\pi/mT}$ is the thermal de Broglie wavelength. More specifically, we have~\cite{Ozawa_2014} 
\begin{align}
\bar s = 2\frac{\calP}{\calD} - x; \quad \bar c_v = 2\frac{\calP}{\calD} - \frac{\calD}{\calD'};\quad \kappa_T = \frac{m}{\rho T} \frac{\calD'}{\calD}; \quad \kappa_s = \frac{m}{\rho T} \frac{\calD}{2\calP},
\end{align}
where $\calD' \equiv d \calD /d x$, $\bar s = ms$ is the entropy per particle  and $\bar c_v = m c_v$  specific heat capacity per particle. Since we are interested in the temperature dependence of many physical quantities, it is convenient to measure the temperature in terms of the BKT transition temperature $T_c$. For a gas with fixed density,  it is easy to see from Eq.~(\ref{calPcalD}) that 
\begin{align}
 \frac{T}{T_c} = \frac{\calD(x_c,g)}{\calD(x,g)},
 \end{align}  
 where $x_c = \frac{\pi}{g}\ln\left( \frac{\xi_\mu}{g}\right) $ with $\xi_\mu = 13.2$ is the value of the $x$ variable at the BKT transition point. Now,
both $\calP$ and $\calD$ can be expressed in terms of a universal function $\theta(x,g)$ as~\cite{Prokof'ev_I, Prokof'ev_II}
\begin{align}
\calD(x,g) = \pi \left [\frac{x}{g} + \theta(x,g) \right ]
\end{align}
and 
\begin{align}
\calP(x,g) = \calP_c + \ln \left( \frac{\xi_\mu}{g}\right) (x-x_c) + \frac{\pi}{2g} (x- x_c)^2 + \frac{\pi g}{2}\left[ \theta^2(x,g) - \theta^2(x_c,g)\right] - g \left[ \theta(x,g) - \theta(x_c,g)\right],
\end{align}
where  $\calP_c $ is the reduced pressure at the critical point and can be calculated by the Hartree-Fock mean-field theory~\citep{Yefsah}. For the temperature range of our interest, the $\theta$ function can be determined analytically by solving the equation~\cite{Prokof'ev_I, Prokof'ev_II}
\begin{align}
\theta(x,g) - \frac{1}{\pi}\ln\theta(x,g) = \frac{1}{g}(x- x_c) + \frac{1}{\pi} \ln(2\xi_\mu).
\end{align}

The static superfluid density $\rho_s(0)$ can be expressed in terms of another dimensionless quantity 
\begin{align}
\calD_s(x,g) = \lambda_T^2 \frac{\rho_s(0)}{m}.
\end{align} This latter is given by 
\begin{align}
\calD_s = 2\pi \theta -1 
\end{align}
for $\frac{x-x_c}{g} > 0.5$ and 
\begin{align}
\frac{4}{\calD_s} + \ln \frac{\calD_s}{4}  = 1+0.61 \frac{x-x_c}{g} 
\end{align}
for $-0.1<\frac{x-x_c}{g} < 0.5$. 
\section{Calculation of the dynamical dielectric constant}
The determination of the dynamical dielectric constant $\e (\omega)$ due to the vortices is crucial to solving Eq.~(\ref{sounds}) for the first and second sound dispersion.  According to the Kosterlitz-Thouless theory, bound vortex pairs of all separations populate in the superfluid below the critical temperature $T_c$. These vortex pairs are similar to the electric dipoles in the 2D plasma in the sense that they can be polarised by the flow of the fluid, resulting in a counterflow which effectively reduces (or renormalises) the bare superfluid density. The renormalised static superfluid density is given by 
\begin{align}
\rho_s = \rho_s^0/ \tilde \e (r = \infty,T),
\label{rhostatic}
\end{align}
where $\tilde \e(r,T)$ is the static, length and temperature dependent dielectric constant describing the polarisability of vortex pairs separated by distance $r$.  As the temperature increases and surpasses  $T_c$, the vortex pairs with largest separations begin to suddenly dissociate into free vortices, i.e., $\tilde \e (r  = \infty, T_c^-)$ is finite while $\tilde \e (r  = \infty, T_c^+)$ diverges. This leads to a precipitous drop of the superfluid density from a finite value to zero at $T_c$ and a breakdown of dissipationless flow for $T > T_c$. 

The above physical picture was later generalised to account for dynamical situations in the so-called dynamical KT theory~\cite{Ambegaokar1,Ambegaokar2,Ambegaokar3}, where $\e(\omega)$ was introduced to characterise the response of bound vortex pairs and free vortices to oscillating velocity fields at frequency $\omega$ in the long wavelength limit.  It can be shown that $\lim_{\omega\rightarrow 0}\e(\omega,T) = \tilde \e(r = \infty,T)$ and thus the static superfluid density in Eq.~(\ref{rhostatic}) is in fact the zero-frequency component of the frequency-dependent superfluid density in Eq.~(\ref{rhosomegas}), i.e., $\rho_s(\omega = 0)$. Importantly, unlike $\tilde \e(r = \infty,T)$ which becomes singular for $T> T_c$, $\e(\omega)$, and thus $\rho_s(\omega)$, is a continuous and smooth function of temperature for any finite frequency. The reason for this is that while the divergence of $\tilde \e(r=\infty,T)$ at $T_c^{+}$ signifies the dissociation of the vortex pairs of the largest size,  $\e(\omega)$ describes the response of the vortex pairs whose size is smaller than the vortex diffusion length $\sqrt{2D/\omega}$. Here $D$ is the diffusion constant of the quantised vortices. For superfluid Helium films, $\e(\omega)$ can be probed by the famous torsional oscillator experiment~\cite{Bishop_I} which verified the dynamical KT theory. 

As we shall see, the necessary ingredients for calculating $\e(\omega)$ are $\tilde \e(r,T)$ and $\xi(T)$, the latter of which is a correlation length  specifying the size of the largest vortex pairs. Both of the quantities can be determined from the famous Kosterlitz-Thouless recursion relations in terms of three microscopic parameters: the bare coupling constant $K_0 = (\hbar/m)^2\rho_s^0/(k_BT)$, the vortex core diameter $a_0$ and the bare vortex fugacity $y_0 = {\rm exp} \{-\lambda K_0\}$~\cite{Kosterlitz_1973}. Averaging the system over vortex pairs with separation smaller than $a_0 e^l$ gives rise to new parameters $K(l)$ and $y(l)$ determined by the recursion relations
\begin{align}
&\frac{d}{dl}[K(l)]^{-1} = 4\pi^3 y^2(l) \nn \\
&\frac{d}{dl}y(l) = [2- \pi K(l)] y(l),
\end{align}
where $K(l=0) = K_0$ and $y(l=0) = y_0$. The scale-dependent static dielectric constant $\tilde \e (r)$ is defined as
\begin{align}
\tilde \e(r) = K_0 / K\left (l = \ln(r/a_0)\right ).
\end{align}
To calculate $\tilde \e(r)$ we need to know $\lambda$ and $K_0$ (or equivalently $\rho_s^0$).  Currently, the best theoretical estimate for the former is $ \lambda = \pi^2/4$~\cite{Nylen}. As for $K_0$ or $\rho_s^0$, we use the recursion relation to infer its value since we have
\begin{align}
K(l=\infty) &=(\hbar/m)^2\rho_s(0)/(k_BT) \nn\\
& = \frac{1}{2\pi}\calD_s.
\end{align}
In other words, once we know $\rho_s(\omega = 0)$ from calculating $\calD_s$, we can deduce $\rho_s^0$ using the recursion relation. 
Next, the correlation length $\xi(T)$ is determined as the length scale at which $y(l)$ becomes comparable to $y_0$, namely
\begin{align}
y(l_\xi) = y_0,
\end{align}
where $l_\xi \equiv \ln(\xi/a_0)$.

In analogy to 2D plasma, the dynamical dielectric constant at temperature $T$ can be written as  
\begin{align}
\e(\omega,T) = \e_b(\omega,T) + i\sigma_f(T)/\omega,
\label{eomegadef}
\end{align}
 where 
$\e_b(\omega,T)$ is the contribution from the bound vortex pairs and $\sigma_f(T)$ is the ``conductivity" due to the free vortices. Via a Fokker-Planck equation approach, the former is given by~\cite{Ambegaokar1,Ambegaokar2,Ambegaokar3}
\begin{align}
\e_b(\omega,T) = 1 + \int_{a_0}^{\xi(T)} dr \frac{d\tilde \e}{dr} \tilde g(r, \omega),
\label{ebound}
\end{align}
where $a_0$ is the vortex core diameter and $\tilde \e(r,T)$ is the static dielectric constant. The correlation length $\xi(T)$ is a length scale specifying the size of the largest vortex pairs and is thus  infinite below $T_c$ and finite above. Here $\tilde g(r, \omega)$ is a response function obeying the following second-order differential equation with respect to $r$
\begin{equation}
r^2\frac{d^2 \tilde g}{dr^2}+\left (3   -\eta \right ) r\frac{d \tilde g}{dr } - \left (-i \frac{\omega}{\omega_0}\frac{r^2}{a^2_0} e^{-2l_D} +  \eta \right )g+ \eta= 0
\label{diffequ}
\end{equation}
where $D$ is the diffusion constant of the vortices and $\eta (r) = 4T_c\tilde \e(\infty,T_c)/[T\tilde \e(r,T)]$. Here we introduce the main parameter of the dynamical KT theory
\begin{align}
l_D \equiv \ln\sqrt{2D/\omega_0 a_0^2},
\end{align} where  $\omega_0 = c_B k_0$ is the typical frequency associated with sound propagation. Here $c_B = \sqrt{g\rho}/m^{3/2}$ is the Bogoliubov sound velocity and $k_0 = \pi/L$ is the wavevector unit for a system of size $L$ along the propagation direction.  In Ref.~\citep{Ambegaokar1,Ambegaokar2,Ambegaokar3}, the following approximate formula was used for $\e_b(\omega)$ 
\begin{align*}
{\rm  Re} \e_b (\omega) &= \tilde \e \left(\sqrt{{14D}/{\omega}} \right ) \nn \\
{\rm  Im} \e_b (\omega) &= \frac{\pi}{4}\sqrt{\frac{14D}{\omega}} \tilde \e' \left(\sqrt{{14D}/{\omega}} \right ),
\end{align*}
where $\tilde\e'(r) \equiv d \tilde \e(r) /dr$. However, in our calculations, we will avoid such approximations and solve Eqs.~(\ref{ebound})-(\ref{diffequ}) exactly. The free vortex contribution is found to be~\cite{Ambegaokar1,Ambegaokar2,Ambegaokar3}
\begin{align}
\sigma_f &= \left (\frac{2\pi \hbar}{m}\right )^2\left (\frac{D\rho_s^0}{k_B T}\right )n_f \nn \\
& = 2\pi^2\omega_0 K_0 e^{2l_D} \frac{a_0^2}{\xi^2}
\end{align}
where $n_f = 1/\xi^2(T)$ is the density of free vortices and $K_0 = (\hbar/m)^2\rho_s^0/(k_BT)$ is the bare coupling constant. Naturally, no free vortex exists below $T_c$,  i.e., $n_f = 0$ for $T< T_c$. 

\begin{figure}[th]
\centering
\includegraphics[width=0.9\linewidth]{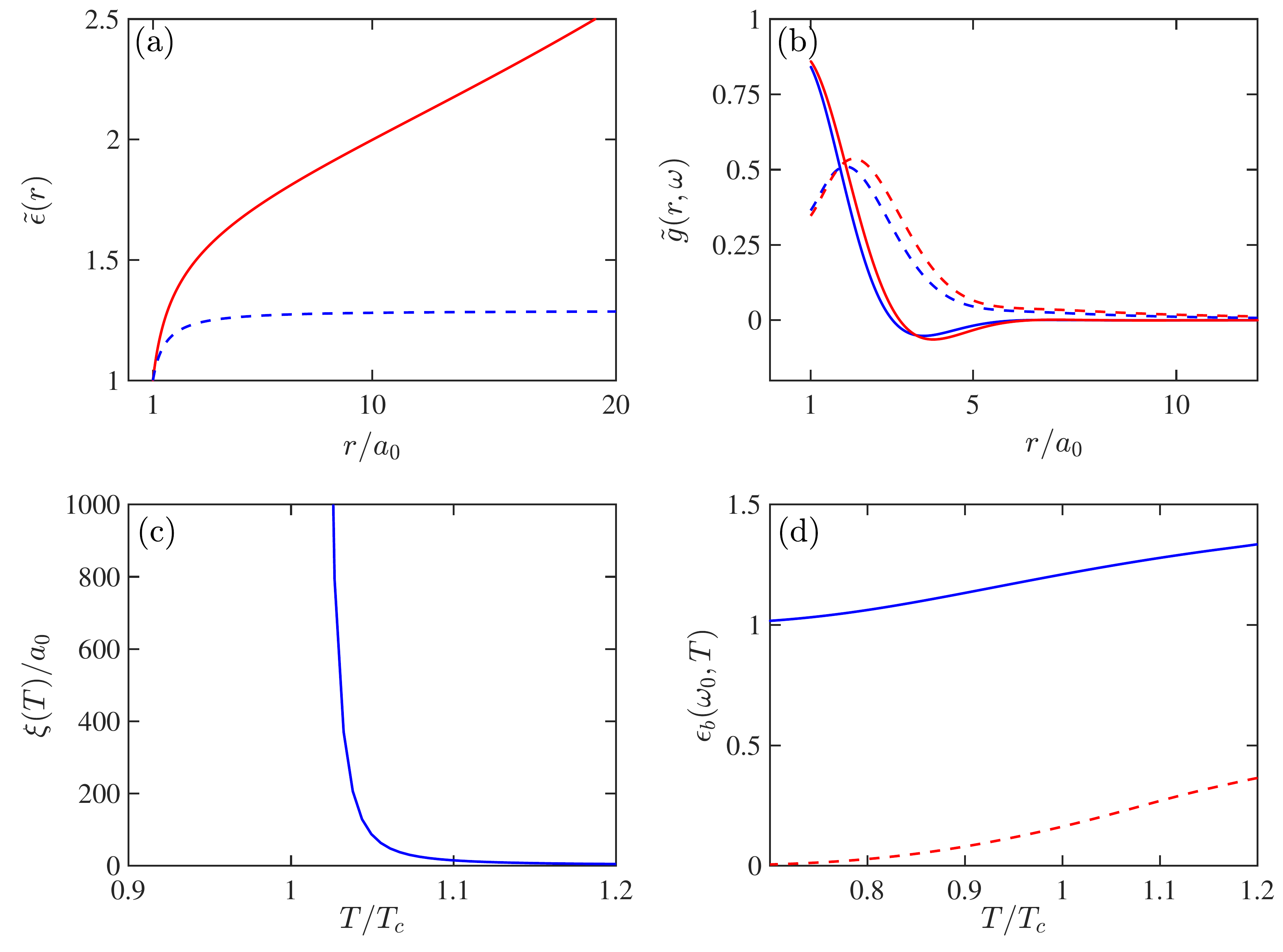}
\caption{(a) $\tilde \e(r)$ calculated for $K_0 = 1$ (solid) and $K_0 = 1.1$ (dashed). (b) Real (solid) and imaginary (dashed) parts of $\tilde g(r)$ calculated $K_0 = 1$ (blue) and $K_0 = 1.1$ (red). (c) $\xi(T)$ as a function of $T/T_c$. (d) Real (solid) and imaginary (dashed) parts of $\e_b(\omega_0,T)$ as a function of $T/T_c$. Here $l_D = -0.5$. }
\label{eb}
\end{figure}

In Fig.~\ref{eb}, we show calculated examples of $\tilde \e(r)$, $\tilde g(r)$, $\xi(T)$ and $\e_b(\omega,T)$ for a weakly interacting 2D Bose gas with the dimensionless coupling constant $g = 0.16$.

\end{document}